\documentclass[twoside
]
{article}
\usepackage{csit}
\usepackage{epsf}
\usepackage{latexsym}
\usepackage{amssymb}
\usepackage{amsthm}
\usepackage{eucal}
\usepackage{amsfonts}
\usepackage{amscd}
\usepackage{times}

\title{\bf Event Driven Objects
}
\author{\large V.E.Wolfengagen\thanks{{\ }Also: Kashirskoye Ave., 31,
Moscow Engineering Physical Institute,
Moscow, 115409 Russia}\\[1.52mm]
\normalsize       Vorotnikovskiy per., 7, bld. 4 \\
\normalsize       Dept. for Advanced Computer Studies and
                  Information Technologies \\
\normalsize        Institute for Contemporary Education ``JurInfoR-MSU''\\
\normalsize        Moscow, 103006 Russia\\
\normalsize        { \tt vew$@$jmsuice.msk.ru}
\date{}}

\institution{}

\newtheorem{exm}{Example}[section]{\rm }{\rm}

\begin{document}
\setcounter{page}{88}

\bibliographystyle{alpha}
\hyphenation{da-ta-ba-ses va-ri-ant}

\markboth{Event Driven Objects}
{Workshop on Computer Science and Information Technologies CSIT'99,
Moscow, Russia, 1999}

\maketitle


\begin{abstract}
\noindent A formal consideration in this paper is given for
the essential notations to characterize the object that is
distinguished in a problem domain.
The distinct object is represented by another idealized
object, which is a schematic element.
When the existence of an element
is significant, then a class of these partial elements is
dropped down into actual, potential and virtual objects.
The potential objects are gathered into the variable
domains which are the extended ranges for unbound variables.
The families of actual objects are shown to be parameterized
with the types and events.
The transitions between events are shown to be driven by the
scripts.
A computational
framework arises which is described by the commutative diagrams.
\smallskip \hrule width 15em  \smallskip
\end{abstract}
%

\section*{Introduction}
\addcontentsline{toc}{section}{Introduction}

The language for a computational environment depends on
the means of data object identification.

The consideration below is given more formal. The essential
notations are summarized to characterize the object that is
distinguished in the problem domain. The term `problem domain' is
enriched to capture the object of current research or investigation.
In particular, the database, mathematical theory, programming system etc.
are the problem domains.

The distinct object is represented by another idealized
object, which is called an element. For convenience this
representation is also called object. The representation is fixed
within the language framework by the formal description. Thus,
the term `element, that represents the generic object' would be
replaced by the term `description-as-object'.

    Furthermore the attention of the researcher is paid to the
analysis of the representation currently mentioned. There is no
need in formal distinction between `object' and `element'. By
need, to resolve the possible misunderstanding the object from the
problem domain and the object as its representation would be
separated (recall that the representation generates the objects
in a mathematical sense). The consideration is not restricted to
a class of the total elements. When the existence of an element
is significant, then a class of the partial elements is covered.
Therefore the possible elements, or objects, are included and this
kind of objects is possible with respect to some a priori
theory.

    Consider an image of some problem domain that contains the
objects and their relationships. This representation becomes
possible when the possibility to distinct the objects and their
relationships does exist. The objects and their relationships are
fixed within some logical language. Suppose that the logical language
contains a set of special individualizing functions. Any its
subset distinguishes an individual. The further consideration is
straightforward: some individualizing functions are truth valued,
thus, preserving the predicativity. Before analysing
the logic thus generated,
a preliminary discussion should give insight to
understand the nature of the object or its representation.

The separation of all the objects into assignments,
individuals (actual, possible and virtual) and concepts
presupposes a special kind of logic with the central notion of an
object or element. Any actual object is implemented by the
appropriate assignment which is applied to the possible object,
and the last one is possible with respect to the prescribed theory.
Under this assumption the elements
exhibit their partial nature.
This means that they are determined and defined with
respect to some subdomains but not with respect to
others.

    Thus, the extended evaluation of the unbound variables is
necessary to represent the partial elements so that they range
over the sets of actual objects (the actual objects which {\em do} exist).
Therefore, the partial objects are the schematic entities and
their schemes are to be considered any time the unbound
variables are evaluated. The schemes give advantage in applying
 theory of computations. A fruitful hint is the following.
The object in a theory of computations is represented by some
scheme that actually transforms the `flow' of information.

For the purely mathematical reasons the term `assignment'
is often used instead of `event' as well as `evolvent' instead of `script'.

The material covered in this paper is as follows.

A general conceptual background is given in Section~\ref{sect:1}.
Its feasibility is based on potentiality, relativity, and separability
of a source set for individuals. The actual individuals are assigned
with the indices while virtual
individuals are added as the idealized entities to increase
the expressivity of a formal language, giving rise to some natural
taxonomy. Some stages of a hypotetical procedure with a problem domain
pre-representation
are selected out as studying of the problem domains,
descriptions with the data, and map to environment.

Section~\ref{sect:2} contains a short outline of the metadata base
issue. Getting started with a global domain $\mathcal H$
the two ways to construe the local domains and concepts are
discussed. The individuals are determined via the functional
entities \mbox{$\langle I, h, T\rangle$} for the further
capturing both the transactioning and cloning effects.
The brief outline, in fact, some extractions, for a type system
are included.

Section~\ref{sect:3} deals more with the semantic considerations.
An approach gives an opportunity to determine the modularity
of persistent couples of data. A primitive frame for the
evaluation map is given by the rules which are the principles
of evaluation.


The basics of variable sets are due to~\cite{Lawv:75}, \cite{Sco:80}.
The reasons for types and structures are influenced, but in a rather
distinct way, by~\cite{Book:91:Asperti:CategoryTheory}.
The semantic considerations for information systems are reviewed
in~\cite{Article:94:Baclawski:CatTheoDatabase}.
Overall and general view for the variety of relations is
influenced by~\cite{Bun:79}.

Other related topics, with an encirclement of computation environment
for different models of information systems,
 are covered in the rest of references.

\section{Conceptual background}\label{sect:1}

The modern trends in developing the majority of information systems
are based on some notion of an object.
A common place is that the meaning of an object
in use depends on the aims and targets of
a particular developer. Typically, the creation of an information system
involves both the database development and implementation stages.
Assume now, that both database and metadata base are under development.
As a rule, not only the data sets but also the metadata couples are appeared
and used as the various universes of concepts.

The metadata objects reflect the rules and restrictions
which are superimposed on the data sets. In addition,
the metadata objects originate an existence of the data objects
and their families. Intuitively, metadata object is observed
as a source to generate these families. Thus, from  a mathematical
point of view metadata object is {\em not} represented by a total
function.

On the other hand a database gives a valid family of the actual
data objects. For the reasons as above this is just a local universe,
or partial domain for metadata.

The feasibility of behaviour for both the sets of
data objects and metadata objects
 meets the difficulties for this partial nature of the domains.
At least, the grounds and rules how to select out, fix and apply the
families of concepts need to be verified with a care. One of the ways
to get started is to revise an imaginable and thus idealized
development procedure.

\subsection{Development grounds}\label{subsect:1-1}

In practice, a database development, even under an adopted data model
and tool, meets a great amount of restrictions and exclusions.
The process seems to be a multistage and iterative,
taking into account the excessive and complicated case study.

An idealization is usually fruitful and needed to concentrate
the efforts on the significant
features and effects and do not lead to an oversimplification.
The formulation of basic development grounds
both for databases and metadata bases needs, at least, the three stages.
Some stages of a hypothetical procedure
are the studying of a problem domain, descriptions with the data,
and map to environment. They are straightforward as follows.

\subsubsection{Studying of the problem domain}

At the first stage, when {\em some problem domain $D$ is studying} the
objects-as-individuals
are selected out. They are treated as the
generic notions which are intuitively clear and transparent, and
 this is safety from a mathematical point of view.

\paragraph{Potentiality}
By the reasons of a formalism the individuals are assumed to be
collected into a single set $\mathcal H$, and this set $\mathcal H$ is
observed as a class of all the {\em potential} individuals.
This assumption seems not to be dramatically restrictive, an important
is namely the {\em possibility} to construe such a single set.
Thus, better way is to fix this set {\em before} the study
to avoid the later contradictions in a target formal
framework. Otherwise, the starting formalism would be revised
under the renewed ideas concerning this initial universe.

\paragraph{Relativity}

The members of this class are evaluated relatively some other
{\em assigned} sets $Asg$ which can be indexed,
so that they can have their own inner structure.
For pure notational reasons to indicate explicitly
 this inner structure, instead of $Asg$, the notation $I$~\label{asg-i}
 will be used.
At this point of discourse
this feature of potential indexing is not necessary but leads to
some reasonable simplifications.

This feature of potentiality reflects the variations of $\mathcal H$
with a time: the individuals {\em enter} some set starting their
existence, and {\em leave} this set cancelling out their existence
relatively previously assigned index. Let observe this index as a cross
reference with the set of all potential individuals.

\paragraph{Separabilty}
To get started with a case study for the behaviour of
individuals, a choice should be done with
{\em actual} $U$, {\em possible} $H$, and {\em virtual} $V$ individuals
which are taken from $\mathcal H$.
The actual individuals are assigned with the indices while virtual
individuals are added as the idealized entities to increase
the expressivity of a formal language.
The natural taxonomy principle reflects all these three sets:
$$
U_i \subseteq H \subseteq V,
$$
where $i$ ranges $Asg$.

The Figure~\ref{fig:01} reflects the
transformations of an in-language individual $\overline{h}$
under evaluation map into the individual $h$, and those
of the individual $h$ under the events $i,j$.
\begin{figure*}
\epsfxsize=2.4in \centerline{\epsfbox{ind-2.ai}}
\caption{\label{fig:01} An evaluation map $\|\cdot\|$ maps the
in-language individual $\overline{h}$ to $h$. This possible
individual $h\in H$ is driven by {\em assignment} (or: index,
event) $i\in Asg$ resulting in the actual individual $h(i)\in
U_i$. The next step could be driven by the event $j\in Asg$ where
the map $h(i)$ results in the state $h(i)(j)$.} \vbox to 1em {\ }
\end{figure*}

\subsubsection{Descriptions with the data}
At the second stage a problem domain is described with the {\em data},
and the truth value of a description becomes sensitive to the
assignment $i\in Asg$, where $Asg$ is a set of all the assignments.
The description is determined by the operator $\mathcal I$,
the prefix ${\mathcal I}x$ is to correspond rather to the words
`the one and only object (entity) $x$ such that'. This description
so determined is a kind of idiom
\mbox{`${\mathcal I}x$ (\dots\ $x$\ \dots)'}, i.e.
$$
\textrm{the object}\ x\ \textrm{such that} \dots\ x\ \dots
$$
is known as {\em description}.
 For instance, the principle
to determine arbitrary individual $h$ is as follows:
$$
\|{\mathcal I}y.\Psi(y)\|i = h \iff
\{h\} = \{\hbar \in H | \|\Psi(\overline{\hbar})\|i = 1 \},
$$
where $\|\cdot\|\cdot$ is an {\em evaluation} function,
$\overline{\hbar}$ is an in-language reference to
the individual $\hbar$ from $H$, $\hbar \in H$. This biconditional means
that the described entity $h$ has a property $\Psi$ under the event
or conditions $i$ and is exactly a {\em sigleton} $\{h\}$. This singleton
is precisely determined by the language construct. This is a
property of singularity for individuals.

\paragraph{Conceptual step}
At the second stage {\em conceptual step}
gives the increase to the degree of a generalization.
This means a transition from the individuals to the
(individual or propositional) concepts.
The value of a concept depends on the reference points
(or assignments) and is a function
(sequence of values which are the individuals).
The principle of a taxonomy for concepts
$$
U^{Asg} \subseteq H^{Asg} \subseteq V^{Asg}
$$
is parallel to those for the individuals.
Some transformations of the concepts  are given
in Figure~\ref{fig:02}.
\begin{figure*}
\epsfxsize=2.4in
 \centerline{\epsfbox{cncpt-2.ai}}
\caption{\label{fig:02}  Let $I\subseteq Asg$ be some subset of
the events. Then a concept $C(I)\subseteq H^I$ generates the
concept $C(\{i\})$ under the event $i\in I$.} \vbox to 1em {\ }
\end{figure*}
The elements of this conceptual taxonomy are the families of the
functions which are not obviously total.

The restricted forms to determine the object $\mathcal H$,
as may be shown, are based on the comprehension principles.
For instance, the inherited comprehension from the higher order logic
gives the natural restrictions to the subclasses by some
properties which correspond to the formulae. Note that this
classification generates a system of types, and this observation
will be used in the further considerations.

Assume, e.g., that so restricted concept $H_T$ is determined by the
{\em comprehension principle} for the sequence of individuals
matching the property $\Phi$:
$$
 H_T \equiv \{h\in H|\Phi\}
         \equiv {\mathcal I}y:[H] \forall h:H(\Phi \leftrightarrow y(h)),
$$
where $[H]$ means the {\em power sort} (power set).
The object $H_T$ above is a subset of $\mathcal H$,
$H_T\subseteq {\mathcal H}$, and type $T$ is determined by the
same equation as $H_T$. The difference between $H_T$ and $T$ is
that $H_T$ have a meaning of a potentially {\em variable} domain
to the contrast to the usual and familiar notion of type $T$.
For  type $T$ we reserve the meaning of a {\em constant} set.
Thus for the individual constants, as they were introduced below in
this subsection on page~\pageref{ind-const},
the difference between $H_T$ and $T$ is not significant.

The concept ${\mathcal H} \supseteq H_T$ so defined determines a behaviour
of the individual sequence, and this is a family with
a parameter $T$.

\paragraph{States}
On the other hand an individual can be observed as the sequence
of {\em states} (or roles) relatively the assignments $i\in Asg$.
This means that the individual $h$ is a function from the set
of assignments $Asg$ to the set of states $S$, $h\in S^{Asg}$,
and $S^{Asg} \subseteq H$. Intuitively, any individual $h\in H$
has its own characteristics relatively assignment $i\in Asg$
so that $h$ is individualized relatively $i$. A set of all these
characteristics determines the individual $h$ in details,
so that $h$ can be understood as the process in a proper
mathematical sense. It means that an individual $h$ corresponds
to the sequence of the states depending on the assignments.

\begin{figure*}
\centerline{\epsfbox{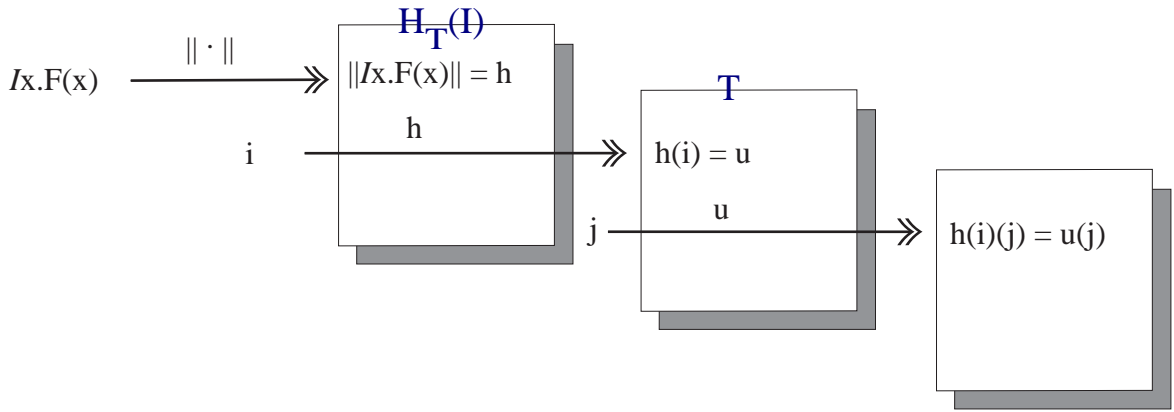}}
\caption{\label{fig:03} The set of individuals \mbox{$h\in H$} generates
the  variable domain \mbox{$H_T(I) = \{h|h: I \to T\}$} whenever
the individuals $u=h(i)$  range a type set $T$.
In addition, note that \mbox{$H_T(\{i\})\subseteq T \subseteq H$}.}
\vbox to 1em {\ }
\end{figure*}

\paragraph{General interconnection}
A general interconnection of the concepts, individuals, and states
can be simulated by the parallel taxonomy:
$$
(S_U^{Asg})^{Asg} \subseteq (S_H^{Asg})^{Asg}
                  \subseteq (S_V^{Asg})^{Asg}
$$
The diagram in Figure~\ref{fig:03} reflects the interrelations
of all the constructs. The stepwise procedure starts with
the in-language description ${\mathcal I}x.F(x)$. Upon evaluation,
it results in the individual $h = \|{\mathcal I}x.F(x)\|$.
When some event $i$ occurs, this leads to $u$ (which is equal to $h(i)$).
The valid subsets $T$ of the individuals $u$ give
rise to the {\em types},
which are in use within the {\em variable domains}
$H_T(I) = \{h|h: I \to T\}$. In case when $h=h(i)\in T$ then we deal with
the individual {\em constants}\label{ind-const}.

\subsubsection{Map to environment}

At the third stage the data objects are mapped to a computational
environment.
In case of the entities above, a host computational environments
must be {\em extensible} to capture all the amount of their
functional nature and behaviour. The reasons to adopt
an event driven model leads to the families of the variable
domains.

\subsection{Existence of elements}\label{subsect:1-2}

In the discussion above all these objects are driven by the events,
and are partial in their nature. Their behaviour needs
a special kind of logic.
The logic of partial objects\footnote{This topic is not included
into this draft.}
 naturally, as may be shown, relativizes the
quantifiers from the greater domains to the subdomains.

\section{(Meta)data base design}\label{sect:2}

A general view to the design procedure drops down to the following
partitions:
\begin{itemize}
\item   study and restrict the problem domain,
\item  determine in a problem domain the data objects' sets. They are
elements and the relations between elements which are described in a
language,
\item  embed the {\em mapped} problem domain (MPD) onto the database (DB),
\item  embed DB into the computational environment,
\item select out the metadata objects, use them as a problem domain
      and repeatedly apply to this domain the steps given above,
\item remap the entities to increase the efficiency of DB.
\end{itemize}
In case the target DB is multilayered the procedure above should be applied
to every layer which is determined by the triples
{$\langle$concept,~individual,~state$\rangle$}.

\begin{figure*}
\hspace{0.15in}\centerline{\epsfbox{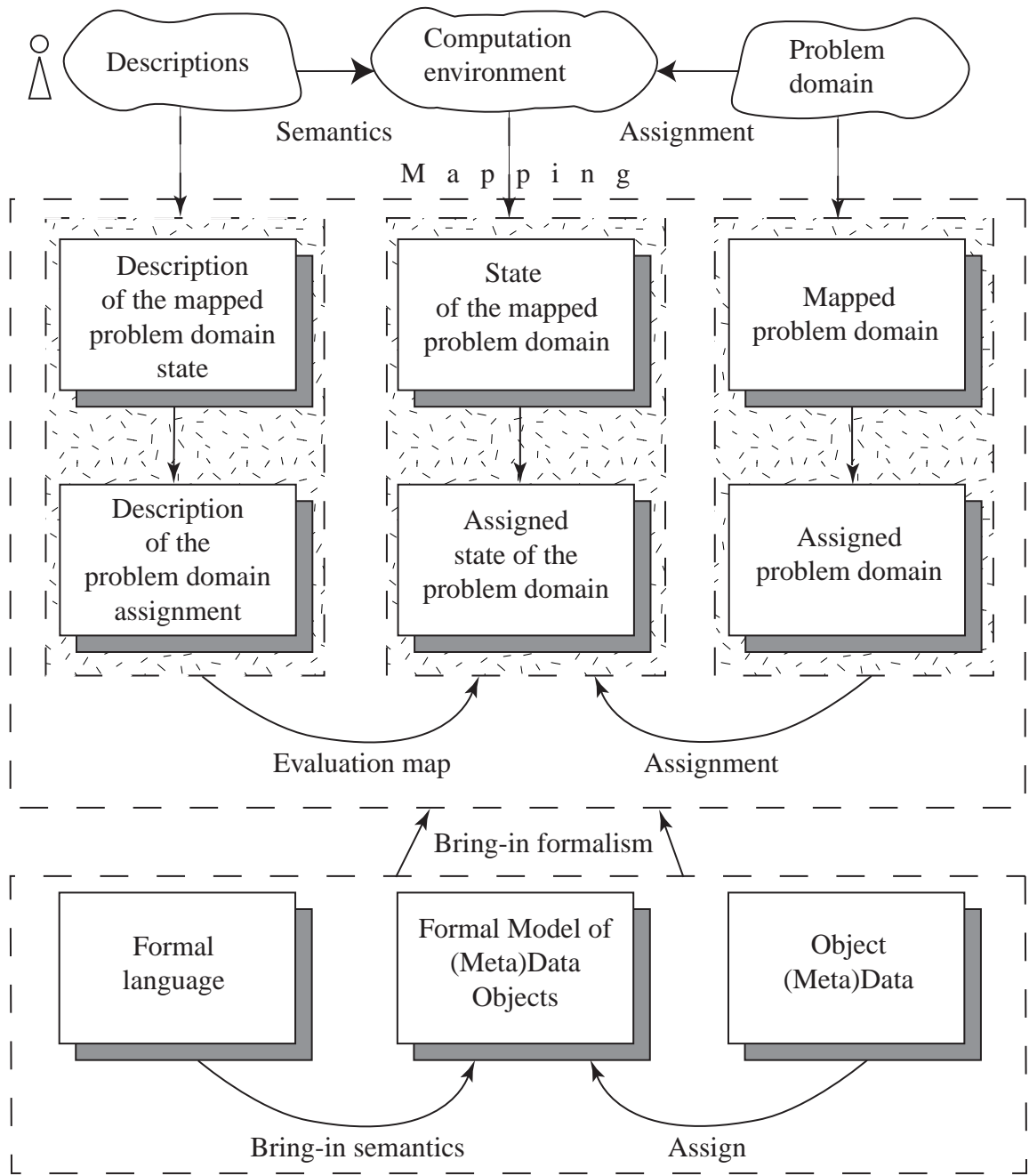}}
\caption{Setting up the assignment of a problem domain. \label{fig:04}
        {\em Explanation}. The descriptions are fixed in a language
        of the user. They determine the state and assignment
        of the mapped problem domain.
        In addition they bring in the semantics
        via some prescribed formalism. On the other hand
        the computation environment captures an assigned
        problem domain.}
\vbox to 1em {\ }
\end{figure*}

Possibly, the most important for success in applying this approach
is an ordering and arrangement of the counterparts which compose
all the supply of ideas in use. In this case a main point is to determine
the process of fixing the reference points for the problem domain.
One of the features is to analyse the {\em process of transition},
whenever the data objects' base, changes the configuration from an {\em old}
state to the {\em new} state. This transition process is viewed as an image
of changes which reflect the changes in a generic problem domain. Thus,
the transitions are viewed as a {\em dynamics} of the problem domain.

In case a data objects' base, which is the state of assigned problem domain,
corresponds to the basic set $M$ in a formalism, this set appears extremely
small. This local nature of the set $M$ becomes too restrictive in comparison
with a universe of all the sets $\mathcal H$ (and $H$).
This univers is continuosly covered step
by step when the state transitions of a database are taken into account.

Indeed, if $M$ and $N$ are the states (sets) then a transition is determined
by the map $f: M\to N$ from the set $M$ into some other set $N$.
In this case the local map $f$ gives no information how to deal with the
operations on all the sets from $H$ into $H$.

The improvement of a `local' formalism upto the `global' one would be
achieved using instead of the sets $M\in H$ the {\em classes}
$P \subseteq H$, which are valid representatives of the families but
not of the elements. In this case a {\em concept} is corresponded to the
class, and every class is represented by the mapping from $H$ to $H$.
Thus, the universe is given by $\mathcal H: H \to H$.

\subsection{Local concepts}\label{sect:2-1}

To exemplify what is a way to construe the concepts for all $I \in H$,
we determine the following alternative sets:
\[\begin{array}{ll}
{\mathcal H}(I) = \{I\}, \\
{\mathcal H}(I) = I \times T & \textrm{where $T$ is a fixed set}
\end{array}
\]
The examples below give more detailed outlook what are the features
and benefits to make a choice.
\begin{exm} {\em First approach to construe the concepts.}\label{exm:01}
Assume now the first alternative,
 i.e. the definition ${\mathcal H}(I) = \{I\}$ is selected out.
\end{exm}
This choice leads to the following way to construe the entities:
\begin{description}
\item the set of reference points, or assignments $I$ is selected out
as an element from the universe $H$ of the individuals, so that $I\in H$;
\item an individual $h$ is represented by the usual map which depends on
the states $i\in I$ (note that $h: H\to H$ !);
\item the maps $h$ determine the concepts as the class generating operators
because \mbox{${\mathcal H}(I) = \{h(i)|i \in I\}$};
\item an operator (or: concept) $\mathcal H$ determines the individuals $h$
because
$$
\hbar = h(i) \leftrightarrow \{\hbar\} = {\mathcal H}(\{i\});
$$
\item hence, a concept $\mathcal H$ is completely determined by the
class introduced above, because
$$
{\mathcal H} = \{[\hbar, I]|\hbar \in H \& I\in H \& \hbar \in {\mathcal
H}(I),
$$
where $[\cdot,\cdot]$ is the operator of an ordered pair.
\end{description}
The observation shows that this construct of metadata object $\mathcal H$
depends on an assignment $I$ to the contrast to the dataobject $h$
which depend on a state $i$.

\begin{exm} {\em Second approach to construe the concepts}\label{exm:02}
Let the second variant is
taken, i.e. the definition ${\mathcal H}(I) = I \times T$ is selected out.
\end{exm}
This way to construe the concepts besides an assinment $I$ involes
the fixed set $T$ which would be better understood in a correspondece
with {\em types}. The following steps reflect an idea:
\begin{description}
\item the assignment $I$ and the set $T$ are fixed and both of them
are the elements of $H$, namely $I\in H$ and $T \in H$;
\item an individual is represented by the map which depends on $i\in I$
and $h: H\to H$ (note that $h(i)\in T$ !);
\item the individuals are generalized to the pairs
$[i,h(i)]$ = $\langle J, h\rangle i$, where $J$ means an identity map,
so that $J\equiv 1_I$,
and $\langle\cdot,\cdot\rangle$ is the coupling function;
\item so generalized individuals $\langle J, h\rangle$ really determine
the concepts because
$$
{\mathcal H}(I) = \{[i, h(i)]|i\in I\} = \{ \langle J, h\rangle i|i\in I\};
$$
\item a concept $\mathcal H$ determines the individuals because
$$
\hbar = \langle J, h\rangle i \leftrightarrow
               \{\hbar\} = {\mathcal H}(\{i\});
$$
\item the concept $\mathcal H$ defined as above is completely determined
by the introduced class because
$$
{\mathcal H} = \{[\hbar, I]|\hbar\in H \& I\in H \& \hbar \in {\mathcal H}(I)
\}
$$
\end{description}

Thus, the metadata object $\mathcal H$ depends on an assignment $I$
but the dataobject $\langle J, h\rangle$ depends on some state $i$.

The comparison of both the examples~\ref{exm:01} and~\ref{exm:02}
leads to some observation. The first kind of choice leads
to a {\em type free} system, where the range $h(i)$ of maps $h$
is the same as the universe $H$. To the contrary, the second approach
to fix the objects leads to a {\em typed system}, where the range $h(i)$
of maps $h$ is restricted by $T$.
Using one of the schemes, the highly dynamical data objects are generated,
they are quite unrestricted by an object taxonomy inherited from
the generic problem domain. Using the other, the more restricted objects
are generated, and the restriction is inherited from the object taxonomy
imposed on the entities within the problem domain. Both the approaches
are not firmly controversial, the resulting concepts are similar,
sharing the same notation as $\mathcal H$.

\subsection{Individuals}\label{sect:2-2}

Usually the development of an information system preserves a
separation of the programming system, descriptions
and operations in a host language, and formalization. This separation
increases the complexity of the tools and decreases the database
dynamics. This is due to the way of data objects' representation
which uses an idea of the constant data objects and their sets.

To the contrast, now the data objects being the individuals
are the dynamical objects and are represented by the functional entities
which are the triples
$$
\langle I, h, T\rangle,
$$
where $I$ is an assignment, $h$ is an individual,
and $T$ is a fixed set, or type. This functional entity is
characterized by
$$
h\subseteq I \times T \& \forall i\in I \exists!\hbar\in T.[i,\hbar]\in h
$$
In case the problem domain is assigned, assume that $\hbar = h(i)$.
An immediate consequence from the characteristics above
is the ability to use the distinct representations, namely,
$[i, hi]$ for $i \in I$, $h(i)\in T$, and $h \subseteq I \times T$
for some individual which is selected out in a problem domain.

\begin{figure*}
\hspace{0.3in}\centerline{\epsfbox{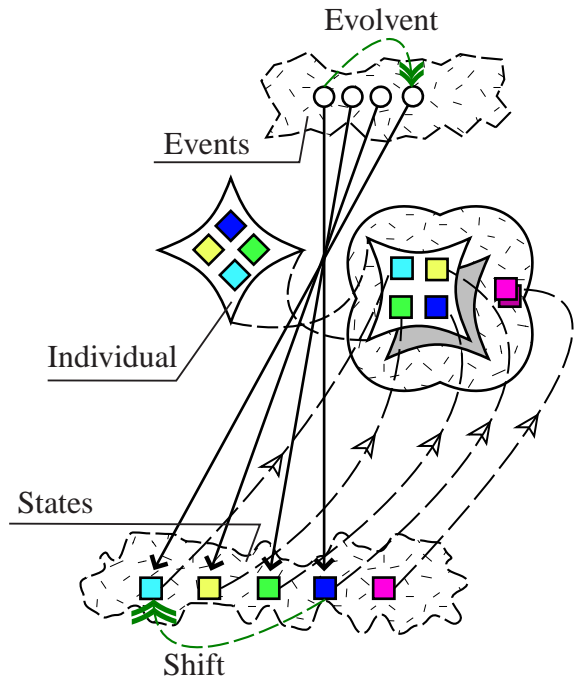}}
\caption{Events, concepts and individuals\label{fig:05}}
\vbox to 1em {\ }
\end{figure*}

An immediate observation shows that the representations
$[i, h(i)]$ and $\langle J, h\rangle i$ are interchangeable
so that
$$
[i, h(i)] = \langle J, h\rangle i
$$
The second expression has a benefit of using
the formal object $\langle J, h\rangle$ which is a map from $i$.
In case the connection with an {\em applicative computation system}
is important then the biconditional below is also valid:
$$
[i, h(i)] \leftrightarrow h \stackrel{\eta}{=} \lambda i.h(i),
$$
where the $\eta$-extensionality from the $\lambda$-calculus is used,
$$
X = \lambda x.X x\ \textrm{whehever}\ x \notin X \eqno{(\eta)}
$$

Up to the moment all the reasons above can be briefly outlined
as follows.
\begin{description}
\item Using the notion of a type, the concept $\mathcal H$ is determined
by $\mathcal H \equiv H_T$,
where the subscript indicates the fixed set $T$.
\item The definition of a class which depends on the assignment $I$
with a fixed set $T$ is given by
$$
H_T(I) = \{h|h: I \to T\}.
$$
\item The object $H_T(I)$ involves the individuals $h$ = $\lambda i.h(i)$
which brings in the extentionality by $(\eta)$-rule from the
$\lambda$-calculus.
\item In addition, the object $H_T(I)$ is determined via the functional
entity $\langle I, h, T\rangle$, and gives the domain $i\in I$
and range $h(i) \in T$.
\item Intuitively, if $I$ is a `state of knowledge', then
$\mathcal H(I)$ is the class of individuals which are known in $I$.
Otherwise, for a fixed set of individuals $T$ the object $H_T(I)$
is the class of individuals, known in $I$ with the {\em property} $T$.
This meaning gives a clear reason to build the system of types.
\end{description}
The ideas and reasons in use are outlined in Figure~\ref{fig:05}.

\subsection{A system of types}\label{sect:2-3}

The objects $H_T(I)$ can be subdivided into subsets $C(I)$
using the natural inclusion $C(I)\subseteq H_T(I)$.
This is to be done via the {\em evaluation map}
$$
\|\Phi(x_1,\dots,x_n)\|\cdot :
I \times (T_1\times \dots \times T_n) \to [\ ],
$$
where the product of types is generated from the types of free variables
$x_1,\dots,x_n$ within the evaluated statement $\Phi$,
and $[\ ]$ is the boolean type. To save writing, this map is indicated
as
$$
\|\Phi(x_1,\dots,x_n)\|\cdot : I \to
               [T_1\times \dots \times T_n]
$$
For a single variable $x$ this leads to the biconditional
$$
\|\Phi(x)\|I  \leftrightarrow C(I)
$$
where
$$
C(I) = \{h\ |\ \|\Phi(\overline{h})\|=1\} \subseteq H_T(I)
$$
To get more familiar with computations using the instances $i$
of the assignments $I$ note that for variable $x$ of type $T$
the map
$$
\|\Phi(x)\| : I \to [T]
$$
ranges over the elements of a powerset $[T]$ for {\em subtypes} of $T$
so that
$$
\|\Phi(x)\|i \in [T],\ \textrm{or}\ \|\Phi(x)\|i \subseteq T
$$
This expression can be rewritten in terms of $C(I)$ as follows:
$$
C(\{i\}) \subseteq H_T(\{i\}) \subseteq T
$$

\section{Semantics}\label{sect:3}

A general semantic consideration leads to the `abstract machine'.

Instead of getting started with the algebraic details the
effort will be applied to some selected topics to illustarte
an approach.

\subsection{Modularity}\label{subsect:3-1}

A notion of the {\em relational module} enables  the relative
independance of a separate \mbox{(meta-)level} from another.
The individuals of a selected out current layer are invariant
from the `lower' level, hence they
do not depend on this lower
and more detailed layer. The `upper' layer is more comprehended
and contains the invariants from the current layer.

The lower layer is generated by the {\em assigning} to the
contrast to the upper layer which is generated by the step
of {\em comprehension}. The comprehension is potentially unrestricted.
The step of comprehension for the current layer $j$
results in the {\em concepts} which in turn are the {\em individuals}
of the neighbouring upper layer $j+1$, where $j \ge 0$:
\[\begin{array}{lcl}
T^{j+1} &\equiv&
\{h^j:\underbrace{[\dots[}_{j\ \textrm {times}} D]\dots] | \Phi^j\} \\
&\equiv & {\mathcal I}y^{j+1}:\underbrace{[\dots[}_{j\ \textrm{times}}D]\dots] \\
&&       \forall h^j : \underbrace{[\dots[}_{j-1\ \textrm{times}}D]\dots]
       (\Phi^j \leftrightarrow y^{j+1}(h^j))
\end{array}\]

\subsubsection{Relative completeness}

A {\em local completeness} of the layer $j$ is presupposed as the
basic property. The layer $0$ gives the ususal relational model.

A {\em completeness} of the model in general is based on a choice
of the appropriate system of assignments. This choice leads to
a {\em family} of images for the problem domains (PD):
\[\begin{array}{lll}
\textrm{layer}\ 0&PD^0 & (\textrm{individuals}) \\
\textrm{layer}\ 1&PD^1 & (\textrm{concepts}) \\
\textrm{layer}\ 2&PD^2 & (\textrm{concepts of concepts}) \\
\textrm{layer}\ 3&PD^3 & (\textrm{concepts of concepts of concepts}) \\
\textrm{layer}\ \dots&\dots & \\
\textrm{layer}\ j&PD^j& (\textrm{concepts of concepts \dots of concepts}) \\
\textrm{layer}\ \dots&\dots &
\end{array}\]

\subsubsection{Intuitive transparency}

Some intuitive reasons are to restrict the comprehension using
the layers $0$, $1$, and $2$.

For layer $0$ the {\em states} correspond to the {\em roles},
{\em concepts} (notions) correspond to the {\em types}, assignments by
the (meta$^0$)events correspond to frames, assignments by the
`(meta$^0$)worlds' correspond to the data base.

For layer $1$ the states are the individuals of the layer $0$,
concepts are the meta$^1$notions, assignments by the meta$^1$events
are the meta$^1$frames, assignments by the `meta$^1worlds$' are
the knowledge base, i.e. the meta$^1$data base.

For layer $2$ the states are the concepts (individuals from $PD^1$),
concepts are the meta$^2$notions, assignments by the meta$^2$events
are the meta$^2$frames, assignments by the meta$^2$worlds are
the metaknowledge base.

To fix the semantics for this model needs to fit the {\em evaluation map}
to the system of assignments.

\subsection{Evaluation map}\label{subsect:3-2}

To match the modularity and intuitive transparecy as above,
the main principles of evaluation the expressions are assumed
as follows.

\subsubsection{Evaluation of application}

The easiest computational idea is due to an {\em applying}
one object, say $M$, to another object, say $N$. Assume, but timely, that
both of the objects are selected out from the same layer
so that $M$ is a function, and $N$ is the argument. The application
$(M N)$ means the value of a function $M$ on the argument $N$.

The assumption is that
{\em the value of application equals an application of the values}
so that the evaluation of the initial expression drops down to
the evaluations of its subparticipants. The above formulation
should be accommodated to both the evaluations with determined
and indetermined assignment.

For an assignment $i$ this gives ({\em Rule~1})
for any $M$, $N$, and as follows:
$$
\left .
\begin{array}{lcl}
\|(M N)\|i &=& (\|M\|i)(\|N\|i)\  \\
\|(M N)\| &=& {\lambda}i.(\|M\|i)(\|N\|i)
\end{array}
\right \}                    \eqno(\textrm{ Rule 1})
$$
\subsubsection{Evaluation of a pair}

A main assumption is that {\em evaluation of a pair equals the
pair of evaluations}. The formalisation drops down to the following
({\em Rule~2}) for any $M$,$N$ and fixed $i$:
$$
\left . \begin{array}{lcl}
\|[M,N]\|i &=& [\|M\|i,\|N\|i]\  \\
\|[M,N]\|   &=& \lambda i.[\|M\|i,\|N\|i] \\
         &\equiv& \langle \|M\|, \|N\| \rangle
        \end{array}
\right \}                     \eqno(\textrm{Rule 2})
$$
Both the principles and derived rules are of basic importance
but lead to the $\lambda$-expressions.

\subsubsection{Evaluation of a $\lambda$-expression}

Assume the application $(\lambda.M)\overline{d}$, where $M$,
$\overline{d}$ are the arbitrary objects, and $\lambda.M$ denotes
an abstraction on {\em some} variable. The derivation of a computational
principle drops down to the following equations:
\[\begin{array}{lcll}
\|(\lambda.M)\overline{d}\|i
      &=& (\|\lambda.M\|i)(\|\overline{d}\|i)
            &  (\textrm{Rule\ 1}) \\
      &=& (\|\lambda.M\|i)d
            &  (\textrm{$(\|\overline{d}\|i) \equiv d$}) \\
      &=& \Lambda \|M\|i d
            &  (\textrm{$\Lambda$ def.}) \\
      &=& \|M\|[i,d]
            &  (\textrm{$\Lambda h \equiv \lambda xy.h[x,y]$})
\end{array}\]
Hence,  for arbitrary $M$ and $\overline{d}$:
$$
\left . \|(\lambda.M)\overline{d}\|i = \|M\|[i,d]\
\right \} \eqno (\textrm{Rule 3})
$$

\subsubsection{Geneartion of $\varepsilon$}

There is an alternative way to evaluate application using
the `dynamical' object $\varepsilon$ which represents the
metaoperator of application. The derivation is as follows:
\[\begin{array}{lcll}
\|(M N)\|i &=& (\|M\|i)(\|N\|i)
    &  (\textrm{by Rule 1}) \\
           &=& \varepsilon[\|M\|i,\|N\|i]
    & (\textrm{by def. $\varepsilon$}) \\
           &=& (\varepsilon \circ \langle \|M\|,\|N\|\rangle)i
    &  (\textrm{by def.
               $\langle\cdot,\cdot\rangle$})
\end{array} \]
Therefore, the following rule is derived:
$$
\left . \|(M N)\| = \varepsilon \circ \langle \|M\|,\|N\|\rangle
\right \} \eqno (\textrm{Rule 4})
$$

\subsubsection{Evaluation of a constant}

A particular assumption deals with the evaluation of a constant:
$$
\left . \begin{array}{lcll}
       \|c\|i &=&c& \textrm{for any $i$} \\
       \|c\|  &=& \lambda i.c
        \end{array}
\right \} \eqno(\textrm{Rule 5})
$$
This means that {\em the constants do not depend on an event},
and this is a principle to be adopted.

\section*{Conclusions}
\addcontentsline{toc}{section}{Conclusions}
A feature analysis of the partial elements
and their corresponding classes was given.
\begin{description}
\item The universe $\mathcal H$ of partial elements, as was shown,
captures the meaning to evaluate the expressions with data objects.
This is a global universe which gives rise to the descriptions
with data. Some distinct local universes are suitable to create
the variable domains.
\item The variable domains are represented by the constructions
which are the families parameterized by the class of events
and the class of types. A variable domain contains the class of
potential elements whose property is identified by the type symbol.
The actual elements are generated by the flow of events.
\item A representation of the class of events, possibly, has the
inner structure. This is important to establish and study the
dynamic features of data model.
\end{description}
As was shown, the universe $\mathcal H$ can be properly separated
to establish the links with the computation models.
\addcontentsline{toc}{section}{References}
\newcommand{\noopsort}[1]{} \newcommand{\printfirst}[2]{#1}
  \newcommand{\singleletter}[1]{#1} \newcommand{\switchargs}[2]{#2#1}

\end{document}